# Self-Motion Mechanism Of Chained Spherical Grains Cells

Sparisoma Viridi[1*] and Nuning Nuraini[2]

[1]*Nuclear Physics and Biophysics Research Division,*
*Institut Teknologi Bandung, Bandung 40132, Indonesia*
[2]*Industrial and Financial Mathematics Research Division,*
*Institut Teknologi Bandung, Bandung 40132, Indonesia*
* Email: dudung@gmail.com

**Abstract.** Cells are modeled with spherical grains connected each other. Each cell can shrink and swell by transporting its fluid content to other connected neighbor while still maintaining its density at constant value. As a spherical part of a cell swells it gains more pressure from its surrounding, while shrink state gains less pressure. Pressure difference between these two or more parts of cell will create motion force for the cell. For simplicity, cell is considered to have same density as its environment fluid and connections between parts of cell are virtually accommodated by a spring force. This model is also limited to 2-d case. Influence of parameters to cell motion will be presented. One grain cell shows no motion, while two and more grains cell can perform a motion.

## INTRODUCTION

Cell motion is an interesting field to discuss. The motion can be addressed to cytoskeletally driven contractions and expansions of the cell membrane [1], drag from fluid medium [2], rigidity of substrate [3], cytoplasmic streaming and membrane protrusions and retractions [4], and myosin-based contractility and transcellular adhesions [5]. Simple mechanism based on only pressure difference is proposed in this work for a 2-d spherical grains based cell.

## CELL MODEL

A cell is modeled by a 2-d spherical grain or connected spherical grains. Each grain has radius $R_i$ at time $t$ and always the same density $\rho$. Cell is located in a fluid with the same density as the cell. It is also assumed that cell motion does not change its depth in fluid, so that hydrostatic pressure suffered by the cell remains the same as $p_0$. Total force per length due to the pressure $p_0$ for a single grain at time $t$ will be

$$F_i = \frac{1}{2} p_0 R_i^2 \int d\theta , \qquad (1)$$

where $\theta$ is angular components in polar coordinate system.

A single spherical cell makes integration of Equation (1) zero. For a cell consisted of two connected grains, there is intersection length $L$ between two grains, which is assumed to have always the same value. The length $L$ will produce a non-zero net force per length for both grains.

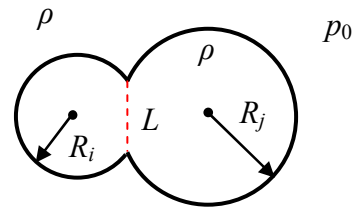

**FIGURE 1.** Cell consists of two connected grains with radius $R_i$ and $R_j$ and also intersection length $L$.

Left grain in Figure 1 will have net force per length $F_i$ to right, while right grain will have net force per length $F_j$ to left. Both this net force per length is due to surrounding pressure $p_0$. During changing radius of $R_i$ and $R_j$ total area of the cell is maintained constant. It means that

$$\frac{d}{dt}\left(\pi R_i^2 + \pi R_j^2 - A_L\right) = 0, \quad (2)$$

where $A_L$ is intersection area. Equation (2) indicates that $R_i$ and $R_j$ depend on each other, so that only one variable can be considered a free variable.

Intersection area can be obtained through equating the two circle equations located at $(x_i, y_i)$ and $(x_j, y_j)$ with condition that

$$(x_i - x_j)^2 + (y_i - y_j)^2 < R_i + R_j, \quad (3)$$

so that the two grains still overlap to each other. From Figure 1 it can be seen that $L$ is a chord of both circles. Intersection area $A_L$ can found from

$$A_L = A_{\text{seg},i} + A_{\text{seg},j}, \quad (4)$$

with $(x_{I1}, y_{I1})$ and $(x_{I2}, y_{I2})$ are the intersection points of the two circle equations. The area in triangle is

$$A_{\text{sec},i} - A_{\text{seg},i} = \frac{1}{2} L \sqrt{R_i^2 - \frac{1}{4} L^2}. \quad (5)$$

Angle $\alpha_i$ between two radii $R_i$ is obtained from

$$\alpha_i = \operatorname{asin}\left(\frac{L}{2R_i}\right). \quad (6)$$

Then,

$$A_{\text{sec},i} = \frac{1}{2} \alpha_i R_i^2. \quad (7)$$

Substituting Equations (6) and (7) into Equation (5) will produce

$$A_{\text{seg},i} = \frac{1}{2} R_i^2 \operatorname{asin}\left(\frac{L}{2R_i}\right) - \frac{1}{2} L \sqrt{R_i^2 - \frac{1}{4} L^2}. \quad (8)$$

Substitution of Equation (8) into Equation (4), and the result into Equation (2) will give the dynamics of the cell. For cell consisted of three or more 2-d spherical grains, formulation in Equation (4) can be generalized into

$$\frac{d}{dt}\left(\sum_{i=1}^{N-1} \frac{1}{2}\pi R_i^2 + \frac{1}{2}\pi R_{i+1}^2 - A_{L,i,i+1}\right) = 0, \quad (9)$$

with $N$ number of 2-d spherical grains.

## SIMULATION

Since one-grain cell is trivial case and the result can be deduced logically, it will not be simulated. Case of two- and three-grain cell will be performed in simulations.

## Case of two-grain cell

Considered there are two circles with identical radius $R_i = R_j = R_0$ that is modulated through time with a function

$$R_i = R_0(1 + \delta \sin \omega t), \quad (10)$$

with $\delta < R_0$. $R_j$ is not free, since it is connected to $R_i$ through

$$\begin{aligned}&\pi R_i^2 - \frac{1}{2} R_i^2 \operatorname{asin}\left(\frac{L}{2R_i}\right) \\&+ \frac{1}{2} L \sqrt{R_i^2 - \frac{1}{4} L^2} \\&+ \pi R_j^2 - \frac{1}{2} R_j^2 \operatorname{asin}\left(\frac{L}{2R_j}\right) \\&+ \frac{1}{2} L \sqrt{R_j^2 - \frac{1}{4} L^2} = c\end{aligned} \quad (11)$$

where $c$ is a constant. Equation (11) can be solved numerically to obtain $R_j$ from known value of $R_i$ from Equation (10).

Net force per length for each grain is calculated using Equation (1) with $\theta \in \left[\frac{1}{2}\alpha, -\frac{1}{2}\alpha\right]$ for the left grain and $\theta \in \left[\pi + \frac{1}{2}\alpha, \pi - \frac{1}{2}\alpha\right]$ for the right grain.

## Case of two-grain cell

For this case, it is actually similar to the previous one, instead that there is additional middle grain which has two $\alpha$, left and right. The relation in Equation (11) is then expanded into

$$\pi R_i^2 - \frac{1}{2}R_i^2 \mathrm{asin}\left(\frac{L}{2R_i}\right) + \frac{1}{2}L\sqrt{R_i^2 - \frac{1}{4}L^2}$$
$$+ \pi R_j^2 - R_j^2 \mathrm{asin}\left(\frac{L}{2R_j}\right) + L\sqrt{R_j^2 - \frac{1}{4}L^2}$$
$$+ \pi R_k^2 - \frac{1}{2}R_k^2 \mathrm{asin}\left(\frac{L}{2R_k}\right)$$
$$+ \frac{1}{2}L\sqrt{R_k^2 - \frac{1}{4}L^2} = c \quad ,(12)$$

Equation (12) will be more difficult to solve than Equation (11) since there is two independent variables, e.g. $R_i$ and $R_j$, if $R_k$ is considered a dependent variable.

## Approximation

In a case that $R_i \ll L$ Equations (11) dan (12) can be simplified through approximation, e.g. Equation (12) can be turned into

$$c \approx \pi R_i^2 + \frac{1}{4}LR_i + \pi R_j^2 + \frac{1}{4}LR_j, \quad (13)$$

and for Equation (12)

$$c \approx \pi R_i^2 + \frac{1}{4}LR_i + \pi R_j^2 + \frac{1}{2}LR_j$$
$$+ \pi R_k^2 + \frac{1}{4}LR_k \quad . \quad (14)$$

## RESULTS AND DICSUSSION

In this section prediction and simulation results are presented and discussed.

## Motions prediction

Motion predictions of cell consisted of one-, two-, and three-linear-grains are given in Figures 2 – 4.

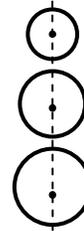

**FIGURE 2.** Motion prediction for cells consists of one 2-d spherical grain: no motion.

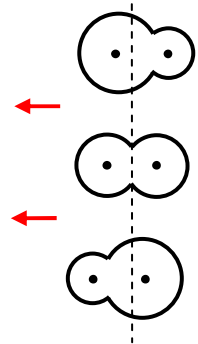

**FIGURE 3.** Motion prediction for cells consists of two 2-d spherical grains: oscillation motion.

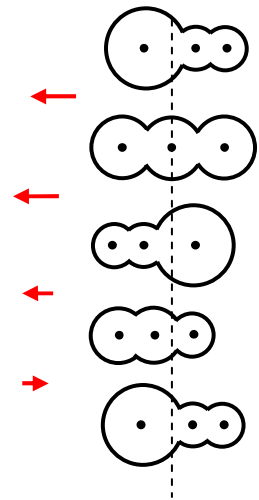

**FIGURE 4.** Motion prediction for cells consists of two 2-d spherical grains: motion in one direction.

One-grain cell can not move since it whatever shrinks or swells net force per length is always zero, as it is illustrated in Figure 2. Two-grain cell exhibits an oscillating motion, that the cell can move forward and backward but not only in one direction. Then, this configuration can not make a motion also, as it is given in Figure 3. Three-grain or more-grain cell can perform a one direction motion, but with cell shrinking and swelling not in the same order and rate. The same ones will only produce the oscillating motion. An example of one direction motion is shown in Figure 4.

## Simulation results

Parameters used in the simulation are given in following Table 1.

**TABLE 1.** Simulation parameters.

| Parameters | Value |
|---|---|
| $R_0$ | 1.0 |
| $\delta$ | 0.2 |
| $\omega$ | 6.2832 |
| $L$ | 0.1 |
| $p_0$ | 1.0 |

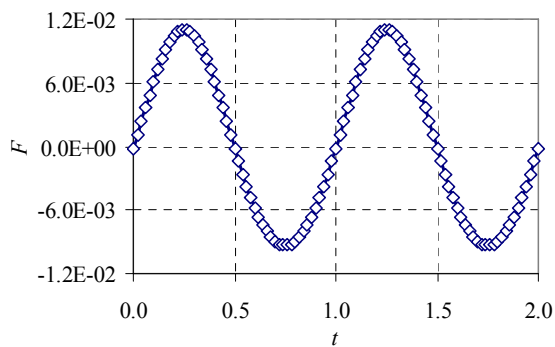

**FIGURE 5.** Net force per length for Case of two-grain cell.

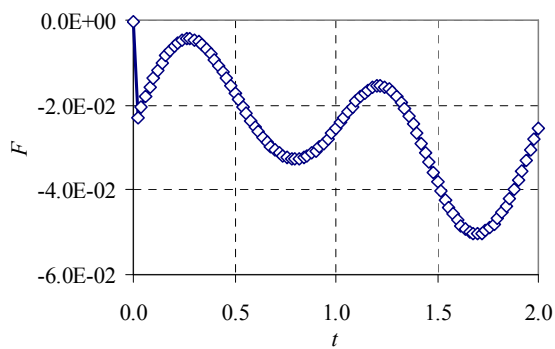

**FIGURE 6.** Net force per length for Case of three-grain cell.

As predicted, Figure 5 shows an oscillating net force per length that will move the cell also in oscillation motion. In Figure 6 it can be seen that the force is not symmetry for positive and negative values, as a matter a fact it always negative, this difference can induce a one direction motion. For this case other $\omega$ is chosen to be half as the previous one. This result shows that the cell will move to left in some backward and forward motion.

It is interesting to investigate non-linier configuration, e.g. circular or triangle grains configuration for more-grain cell, that will be conducted in the next work.

## CONCLUSION

Motion of cell consisted of connected grains has been simulated. One-grain cell shows no motion, two-grain cell performs oscillation motion, and three-grain cell exhibits one-direction motion. More-grain cell and not-linear combinations are subject for next investigation.

## ACKNOWLEDGMENTS

Riset Inovasi Kelompok Keahlian Institut Teknologi Bandung (RIK-ITB) in year 2013 with contract number 248/I.1.C01/PL/2013 supports this work partially.